# Variation of Mass with Velocity: "Kugeltheorie" or "Relativtheorie"

Galina Weinstein

This paper deals with four topics: The first subject is Abraham's spherical electron, Lorentz's contracted electron and Bücherer's electron. The second topic is Einstein's 1905 relativity theory of the motion of an electron. Einstein obtained expressions for the longitudinal and transverse masses of the electron using the principle of relativity and that of the constancy of the velocity of light. The third topic is Einstein's reply to Ehrenfest's query. Einstein's above solution appeared to Ehrenfest very similar to Lorentz's one: a deformed electron. Einstein commented on Ehrenfest's paper and characterized his work as a theory of principle and reasoned that beyond kinematics, the 1905 heuristic relativity principle could offer new connections between non-kinematical concepts. The final topic is Kaufmann's experiments. Kaufmann concluded that his measuring procedures were not compatible with the hypothesis posited by Lorentz and Einstein. However, unlike Ehrenfest, he gave the first clear account of the basic theoretical difference between Lorentz's and Einstein's views. Finally, Bücherer conducted experiments that confirmed Lorentz's and Einstein's models; Max Born analyzed the problem of a rigid body and showed the existence of a limited class of rigid motions, and concluded, "The main result was a confirmation of Lorentz's formula".

## 1. The Law of Variation of Mass with Velocity

### 1.1 Abraham's spherical electron

According to Max Abraham inertia was created by the electromagnetic field in the ether. Therefore, in order to obtain the law of variation of mass with velocity, one had first to calculate the electromagnetic momentum from the electron's self field.

Using the usual definition of the second law of Newton, in the quasi-stationary approximation, Abraham replaced the electromagnetic momentum in the second law of Newton, and converted it into the definition: acceleration times mass equals force. He then obtained the law for the variation of mass with velocity, while assuming rigid spherical electrons, keeping their spherical form at any velocity.

Abraham assumed that the total apparent mass of the electron is not the same when the actual force applied to the electron is parallel with its velocity and tends to accelerate its motion, as when it is perpendicular to the velocity and tends to alter its direction:[1]

$$M_l = \frac{3}{4} \cdot M_0 \frac{1}{\gamma^2} \left\{ \frac{2}{1-\gamma^2} - \frac{1}{\gamma} \cdot \ln\left(\frac{1+\gamma}{1-\gamma}\right) \right\}$$

---

[1] Abraham, Max, "Prinzipien der Dynamik des Elektrons", *Annalen der Physik* 10 1903, pp. 105–179; p. 152.

$$M_t = \frac{3}{4} \cdot M_0 \frac{1}{\gamma^2} \left\{ \frac{1+\gamma^2}{2\gamma} \cdot \ln\left(\frac{1+\gamma}{1-\gamma}\right) - 1 \right\}.$$

Accordingly one distinguishes between the longitudinal mass $M_l$ and the transverse mass $M_t$, and:

$$M_0 = \frac{2}{3}\frac{e^2}{r}, \qquad \gamma = \frac{v}{c}.$$

**1.2 Lorentz's deformed electron**

In 1904, in his paper, "Electromagnetic Phenomena in a System Moving with any Velocity Smaller than that of Light", Lorentz calculated the electron's mass from its electromagnetic momentum too. However, Lorentz assumed that the moving electron was contracted in the direction of motion. He obtained the following equations for longitudinal mass and the transverse mass.[2]

$$M_l = \frac{e^2}{6\pi c^2 r}\beta^3, \qquad M_t = \frac{e^2}{6\pi c^2 r}\beta$$

where $r$ is the electron's radius and,

$$\beta = \frac{1}{\sqrt{1-\frac{v^2}{c^2}}} \text{ and } M_0 = \frac{e^2}{6\pi c^2 r}.$$

Max Abraham had opposed Lorentz's expression, because Lorentz's electron would need to rely upon non-electrical internal forces to sustain it. Still this did not satisfy Abraham's desire for a wholly electromagnetic basis for dynamics. Abraham admitted that the new expression was much simpler than his from a mathematical point of view, but simplicity was not yet a reason for preferring a mathematical expression unless the simple expression was confirmed through research. Abraham adhered to a rigid electromagnetic foundation for a theory of the electron, and also chose a rigid spherical electron.

There was a third law for the variation of mass with velocity proposed by Alfred Bücherer in 1904.[3] This law, like Abraham's, was also compatible with a completely electromagnetic electron theory:

$$M_l = M_0(1-\gamma^2)^{-4/3}, \quad M_t = M_0(1-\gamma^2)^{-1/3}.$$

---

[2] Lorentz, Hendrik Antoon, "Electromagnetic Phenomena in a System Moving with any Velocity Smaller than that of Light", *Verslagen Konignklijke Akademie Van Wetenschapen (Amsterdam). Proceedings of the section of science,* 6, 1904, pp. 809-836; reprinted in Lorentz, Hendrik Antoon, *Collected Papers 1935-1939*, The Hague: Nijhoff, 9 Vols; Vol. 5, pp. 172-197; pp. 184-185.

[3] Bücherer, Alfred, *Mathematische Einführung in die Elektronentheorie*, 1904, Leipzig: B.G. Teubner, pp. 57-58. Langevin had independently proposed the same hypothesis as Bücherer about the shape of a moving electron: Langevin, Paul, "La Physique des electrons", *Review générale des sciences pures et appliquées* 16, 1905, pp. 257-276.

## 2 Einstein's Theory of the Motion of an Electron

### 2.1 Einstein's 1905 expressions for the mass of the electron

Einstein's last derivation in section §10 of the 1905 relativity paper concerned the dynamics of a (slowly accelerated) electron.[4] Einstein obtained expressions for the longitudinal and transverse masses, using the principle of relativity and that of the constancy of the velocity of light.

Consider a particle in motion with a charge *e* (Einstein calls it an "electron"), in an external electromagnetic field. For its law of motion, we assume: **F** = mass x acceleration. If the electron is at rest at a particular instant, its motion during the next instant of time will be according to:

$$m \frac{d^2 \vec{r}}{dt} = e\vec{E}$$

where *m* is the electron's mass, *e* is its charge, *x*, *y*, *z* are its coordinates (*r*) and *t* its time relative to the system *K*.

Einstein assumes that at the moment we are observing the electron, it is at the origin of the coordinates and is moving with velocity *v* along the *x* axis of system *K*. Therefore, at the given moment *t* = 0, the electron is at rest relative to another system of coordinates *k*, which is moving in parallel motion with velocity *v* along the *x* axis of *K*.

Einstein is guided by **the principle of relativity**: the electron is instantaneously at rest in *k*, and the same definition of Newton's second law (*F*= mass x acceleration) is valid according to the principle of relativity for both *K* and *k*. The equations of motion relative to *k* are thus:

$$M_l \frac{d^2 x'}{dt'^2} = eE'_x$$
$$M_t \frac{d^2 y'}{dt'^2} = eE'_y$$
$$M_t \frac{d^2 z'}{dt'^2} = eE'_z$$

(primed terms refer to *k*). Using the Lorentz transformations for coordinates and time and the transformations for the electric and magnetic fields, and applying them to *K*, one obtains:[5]

---

[4] Einstein, Albert, "Zur Elektrodynamik bewegter Körper, *Annalen der Physik* 17, 1, 1905, pp. 891-921; pp. 917-919.
[5] Einstein, 1905, pp. 919.

$$\frac{d^2x}{dt^2} = \frac{e}{m}\frac{1}{\beta^3}E_x$$

$$\frac{d^2y}{dt^2} = \frac{e}{m}\frac{1}{\beta}\left(E_y - \frac{v}{c}B_z\right)$$

$$\frac{d^2z}{dt^2} = \frac{e}{m}\frac{1}{\beta}\left(E_z - \frac{v}{c}B_y\right)$$

Einstein then takes into account "the conventional approach" and inquires as to "the 'longitudinal' and the 'transverse' mass of the moving electron". Using the above equations written in the following form:

$$m\beta^3\frac{d^2x}{dt^2} = eE_x = eE'_x = M_l\frac{d^2x'}{dt'^2}$$

$$m\beta\frac{d^2y}{dt^2} = e\left(E_y - \frac{v}{c}B_z\right) = eE'_y = M_t\frac{d^2y'}{dt'^2}$$

$$m\beta\frac{d^2z}{dt^2} = e\left(E_z - \frac{v}{c}B_y\right) = eE'_z = M_t\frac{d^2z'}{dt'^2}$$

He then notes that $eE'_x$, $eE'_y$, $eE'_z$ are the components of pondermotive force acting on the electron, as viewed in the moving system, moving at this moment with the same velocity as the electron.

Einstein then says, "the force acting on the electron is called", and maintains the Newtonian equation: mass times acceleration = force, while acceleration is measured in the system *K*, and writes the longitudinal mass:

$$M_l = \frac{M}{\left(\sqrt{1-\frac{v^2}{c^2}}\right)^3}$$

and transverse mass:

$$M_t = \frac{M}{\sqrt{1-\frac{v^2}{c^2}}}.$$

In 1905 Einstein took the ordinary point of view and wrote the 'longitudinal' and the 'transverse' mass of the moving electron.

In a 1913 reprint Einstein appended a note to the above word "called": "The definition of force given here [mass times acceleration = force] is not advantageous as was first noted by M. Planck. It is instead appropriate to define force in such a way that the laws of momentum and conservation of energy take the simplest form".[6]

## 2.2 Relativistic Dynamics

Although apparently the 1905 paper dealt with kinematics and electrodynamics, the scope of the paper was beyond these fields and led to the inauguration of relativistic dynamics. Einstein remarked that his results are also "valid for ponderable material points, because a ponderable material point can be made into an electron (in our sense of the word) by adding to it an *arbitrarily small* electric charge".[7] For Einstein ponderable mater is uncharged matter; Einstein was declaring that the ether was superfluous, but he still used the conventional "ether-based" notions. This was the language he could communicate with his colleagues (as much as he could find them while writing the relativity paper as a patent clerk).

But there is much more to Einstein's last conclusion; Einstein tried to explain to his readers something very important: his results concerning the electron (the mass of the electron and the pondermotive force acting on the electron) are also valid for material points. Einstein was thus not only presenting another model for the electron; Einstein in section §10 laid down the starting point for a dynamics of a slowly moving material point – although he himself did not develop a relativistic dynamics.

In 1906 Max Planck inaugurated relativistic dynamics, although Planck still remained within the confines of electrodynamics. Planck defined the second law of Newton, in terms of the rate of change of a new relativistic momentum, in his paper, "The principle of Relativity and the Equations of Mechanics".[8] He extended Einstein's procedure of 1905 in section §10 in the following way:

Planck imagined a mass point, a particle, to be at the origin of the coordinate system *x, y, z, t* (equivalent to Einstein's *K*), and having velocity components, $v_x$, $v_y$, $v_z$ and he required as for the equations of motion of the particle.

Planck considered a new reference system (equivalent to Einstein's *k*), moving with the same velocity as the mass point, relative to *K*; that is, moving along the *x* axis of the original reference system with a constant velocity v, the components of which are, $v_x$, $v_y$, $v_z$. Planck said that the particle moves with finite velocity of size q and parallel to *x*. The particle moves according to the regular Newtonian equation of motion for a free mass point (mass times acceleration = force).

---

[6] *The Collected Papers of Albert Einstein, Vol. 2: The Swiss Years: Writings, 1900–1909* (*CPAE*, Vol. 2), Stachel, John, Cassidy, David C., and Schulmann, Robert (eds.), Princeton: Princeton University Press, 1989, note 41, pp. 309-310. I thank Prof. John Stachel for this.
[7] Einstein, 1905, p. 919.
[8] Planck, Max (1906a), "Das Prinzip der Relativität und die Grundgleichungen der Mechanik", *Verhandlungen der Deutschen Physikalischen Gesellschaft* 4, 1906, pp.136-141, in Planck, Max, *Physikalische Abhandlungen und Vorträge*, 1958, Braunschweig: Frieder. Vieweg & Sohn, Band II, pp. 115-120; p. 118.

However, now Planck defined the force as the electromagnetic force, and thus the mass point moves under the action of an electromagnetic field. He transformed the Newtonian equations of motion to another reference system whose *x*-axis coincides with the direction of the velocity q, that is, relative to which the mass point moves with the velocity *q* relative to *k*. This is actually the system *x, y, z, t* at rest (*K*), but we can call this system, system *k'*. Planck inserted the velocity of the particle *q* instead of *v* into Einstein's transformation for the electric and magnetic fields, and wrote for the system *x, y, z, t* the following equation:

$ma_x/\sqrt{(1 - q^2/c^2)} = eE'_x - (ev_x/c^2)(v_xE'_x + v_yE'_y + v_zE'_z)$

$+ (e/c)(v_yB'_z + v_zB'_y)$, and so on

Planck said that this equation confirms a general result for Einstein's relations (transformation for the electric and magnetic fields), and for the Newtonian equation of motion for a free mass point, for any value of *v*. "We will now bring the equations of motion to their simplest form", said Planck.[9]

Planck multiplied the Newtonian equations of motion with respect to *k'* by $v_x, v_y, v_z$:

$(v_xE'_x + v_yE'_y + v_zE'_z) = m(v_xa_x + v_ya_y + v_za_z)/(1 - q^2/c^2)^{3/2}$,

and inserted this equation into the above equations. In addition, he defined,

$eE'_x + (e/c)(v_yB'_z + v_zB'_y) = F_x$, and so on.

All this led Planck to the following equations:

$d/dt \cdot \{mv_x/\sqrt{(1 - q^2/c^2)}\} = F_x$, and so on,

where these hold for a unit charged mass point moving in an electromagnetic field in *k'*. If *q* is small compared with the velocity of light *c*, these equations reduce to the Newtonian equations of motion. The term inside the brackets is Planck's relativistic momentum.[10] Therefore, Planck defined the second law of Newton as the rate of change of momentum in order for the principle of relativity and the Lorentz transformations to hold good for both systems *k* and *k'*.

In 1907 Einstein adopted Planck's relativistic momentum:[11]

$\xi = \mu v/\sqrt{(1 - v^2/c^2)}$,

and said that Planck's above equations of motion "do not have a physical meaning, but are rather to be understood as defining equations of the force".[12]

---

[9] Planck, 1906a, in Planck, 1958, p. 118.
[10] Planck, 1906a, in Planck, 1958, p. 119.
[11] Einstein, Albert (1907a), "Über das Relativitätsprinzip und die aus demselben gezogenen Folgerungen", *Jahrbuch der Radioaktivität* 4, 1907, pp. 411-462; p. 435.
[12] Einstein, 1907a, pp. 433-434

Planck's equations later led to a single expression for the mass:

$m = m_0/\sqrt{(1 - v^2/c^2)}$.

Although mass could be split along the direction of motion and normally to it, scientists thought it would be conceptually preferable to look upon mass as one quantity. Einstein did not obtain a single expression for the mass, neither in 1905, or later. In later years, at least, Einstein did not talk about a variation of mass with velocity, but only of the new definition of momentum and of energy. Einstein might have rejected or disliked the concept of a single expression for the variation of mass with velocity and therefore did not obtain such an expression.

## 3. The Principles of Relativity as Heuristic Principles

### 3.1 Einstein's reply to Ehrenfest

Einstein had a friend, Paul Ehrenfest a Jewish physicist from Vienna. In 1907 Ehrenfest wrote a paper.[13] There were the known problems in the 19th century electrodynamics of moving bodies. Einstein's 1905 solution appeared to Ehrenfest very similar to Lorentz's solution to these problems: a deformed electron. Ehrenfest thought that Einstein's theory of the motion of an electron could have been obtained from the good old theory of Lorentz, if we only used the method of deduction. If this was so, Ehrenfest understood that Einstein's theory was nothing but a reformulation of the electrodynamics of Lorentz. Therefore, Einstein's innovation was the following according to Ehrenfest, "In the formulation in which Mr. Einstein published it, Lorentzian relativity electrodynamics is treated rather generally as a closed system."[14]

Einstein commented on Ehrenfest's paper. His 1907 reply, "Comments on the Note of Mr. Paul Ehrenfest" is important for the demarcation between his theory of relativity and Lorentz's ether-based theory. Lorentz's theory and the descendants of Lorentz's theory are not theories of relativity. Einstein characterized his work what would be later called principle of relativity as a theory of principle and reasoned that beyond kinematics, the 1905 heuristic relativity principle could offer new connections between non-kinematical concepts, "The principle of relativity, or more exactly, the principle of relativity together with the principle of the constancy of the velocity of light, is not to be conceived as a 'closed system', in fact, not as a system at all, but merely as a heuristic principle which, when considered by itself, contains only statements about rigid bodies, clocks, and light signals. The theory of relativity provides something additional only in that it requires relations between otherwise seemingly unrelated regularities".[15]

---

[13] Ehrenfest, Paul, "Die Translation deformierbarer Elektronen und der Flächensatz", *Annalen der Physik* 23, 1907, pp. 204-205.
[14] Einstein, Albert (1907b), "Bemerkungen zu der Notiz von Hrn. Paul Ehrenfest: 'Die Translation deformierbarer Elektronen und der Flächensatz' ", *Annalen der Physik* 23, 1907, pp. 206-208; p. 206.
[15] Einstein, 1907b, pp. 206-207.

In his 1916 popular book, *Relativity, the Special and the General Theory*, in the chapter "The Heuristic Value of The Theory of Relativity", Einstein wrote: "[…] the theory becomes a valuable heuristic aid in the search for general laws of nature".[16]

Eherenfest's query dealt with the structure of the electron: "Accordingly, it [Lorentz's theory in Einstein's formulation] must also be able to provide purely deductively an answer to the question posed by transferring Abraham's problem from the rigid electron to the deformable one […]".[17]

Einstein answered Ehrenfest's query by saying that the theory of the motion of an electron is obtained as follows: "one postulates the Maxwell equations for vacuum for space-time coordinate systems. By applying the space-time transformation [Lorentz transformation] derived by means of the system of relativity, one finds the transformation equations for electric and magnetic forces. Using the latter, and applying the space-time transformation, one arrives at the law for the acceleration of an electron moving at arbitrary speed from the law for the acceleration of a slowly moving electron (which is assumed or obtained from experience)".[18]

Einstein explained to Ehrenfest, "We are not dealing here at all with a 'system' in which the individual laws are implicitly contained and from which they can be found by deduction alone, but only with a principle that (similarly to the second law of the thermodynamics permits the relation of certain laws to others".[19]

In 1949 Einstein explained this still further: "Gradually I despaired of the possibility of discovering the true laws by means of constructive efforts based on known facts. The longer and the more desperately I tried, the more I came to the conviction that only the discovery of a universal formal principle could lead us to assured results. The example I saw before me was thermodynamics. The general principle was there given in the theorem […the second law of thermodynamics]. How, then could such a universal principle be found?"[20]

### 3.2 Theories of Principle and Constructive Theories

After 1907 Einstein made a distinction between theories of principle, such as thermodynamics and constructive theories, such as statistical mechanics. He characterized the special theory of relativity as a theory of principle, and considered it to be basically complete when the two underlying principles of the theory (the principle of relativity and that of the constancy of velocity of light) were established.

---

[16] Einstein, Albert, *Uber die Spezielle und die Allgemeine Relativitätstheorie*, *Gemeinverständlich*, 1920, Braunschweig: Vieweg Sohn, p. 29
[17] ("ein heuristisches Prinzip"). Einstein, 1907b, p. 206.
[18] Einstein, 1907b, p. 207.
[19] Einstein, 1907b, p. 207. It was the first time that Einstein compared the relativity principle to the laws of thermodynamics. *CPAE*, Vol. 2, p. 412, note 8.
[20] Einstein, Albert, "*Autobiographisches*"/"Autobiographical notes" In Schilpp, Paul Arthur (ed.), *Albert Einstein: Philosopher-Scientist*, 1949, La Salle, IL: Open Court, pp. 1–95; pp. 48-49.

All later work would involve development of constructive theories compatible with these basic principles.

In his paper, "What is the Theory of Relativity?", written at the request of the *London Times* and published on November 28, 1919, for the first time Einstein formulated his views in a systematic manner:[21]

"We can distinguish various kinds of theories in physics. Most of them are constructive. They attempt to build up a picture of the more complex phenomena out of the materials of a relatively simple formal scheme from which they start out. Thus the kinetic theory of gases seeks to reduce mechanical, thermal, and diffusional processes to the movements of molecules – i.e., to build them up out of the hypothesis of molecular motion. When we say that we have succeeded in understanding a group of natural processes, we invariably mean that a constructive theory has been found which covers the processes in question.

Along with this most important class of theories there exists a second, which I will call 'principle theories'. […]

The advantages of the constructive theory are completeness, adaptability, and clearness; those of the principle theory are logical perfection and security of the foundations.

The theory of relativity belongs to the latter class. In order to grasp its nature, one needs first of all to become acquainted with the principles on which it is based".

## 4. Kaufmann's Experiments: "Kugeltheorie" and "Relativtheorie"

**4.1 Kaufmann's Experiments**

In 1906 Planck wrote a letter to Einstein and mentioned his paper "The principle of Relativity and the Equations of Mechanics".[22] Planck told Einstein: "Herr Bücherer has already written to me a letter about his sharp opposition to my latest research, (without giving a reason, to be sure) he declares the principle of relativity incompatible with the principle of least action. It is therefore all the more gratifying to me to see from your card that, for the present, you do not share the views of Herr. B. So long as the proponents of the principle of relativity constitute such a modest group as they do at present, it is doubly important that they agree among themselves […]."[23]

---

[21] Einstein, Albert, *Ideas and Opinions*, 1954, New Jersey: Crown publishers, p. 228; *The London Times*, November 28, 1919. It should be borne in mind that Einstein wrote this article *after* developing the General Theory of Relativity, and when he spoke about the theory of relativity and the principle theory he probably meant both special and general relativity, because he did not write explicitly the word "special".
[22] Planck, 1906a, in Planck, 1958, p. 118 (see section 2.2 above).
[23] Planck to Einstein, July 6, 1907, *The Collected Papers of Albert Einstein, Vol. 5: The Swiss Years: Correspondence, 1902–1914* (*CPAE*, Vol. 5), Klein, Martin J., Kox, A.J., and Schulmann, Robert (eds.), Princeton: Princeton University Press, 1993, Doc. 47.

Planck defended the relativity principle when Walter Kaufmann performed his experiments, the result of which seemed to contradict Einstein's model of the electron.

1900 onward heralded an especially interesting period of experimental researches, conducted by Kaufmann that repeatedly confirmed Abraham's theory, at least until 1905. These and other experiments led to the discovery of cathode corpuscles in the beta radiation of radium, corpuscles that emitted and moved in velocities close to that of light. The beta particles' velocity was substantially faster than that of ordinary cathode rays. Scientists understood that these velocities were fast enough in order to ascertain whether, and to what extent, the inertia of these particles indeed changes with velocity. In 1901 and 1902 Kaufmann's experiments led to the mathematical expression Abraham had predicted, with the aid of some radium chloride, which the Curies had given to Kaufmann. This confirmed Abraham's prediction, according to which the particles owed all their energy to the fact that they were electrified.

Kaufmann wrote in 1902 that the mass of electrons was dependent on velocity; and this dependence could be described exactly by the formula of Abraham. Therefore, the mass of the electrons had a pure electromagnetic character: "The mass of electrons in Becquerel rays depends on the velocity; the dependence is exactly represented by Abraham's formula. The mass of electrons is accordingly of purely electromagnetic nature".[24]

In 1905 Kaufmann concluded that his results: "[...] speak against the correctness of Lorentz's, and also consequently of Einstein's fundamental hypothesis. If one considers this hypothesis as thereby refuted, then the attempt to base the whole of physics, including electrodynamics and optics, upon the principle of relative movement is also a failure".[25]

In 1906 Kaufmann concluded his paper, "On the Constitution of the Electron" by stating that Abraham had proven that the Lorentzian electron required the concept of work, therefore, in order to avoid a conflict with the energy law it was necessary to assume the existence of an "internal potential energy" of the electron. In contrast, the existence of a pure electromagnetic basis for the mechanics of the electrons that would apply to mechanics as a whole would be proven as being impossible.[26] This was so even if one contemplated the existence of a universal external pressure (as Henri Poincaré did in his theory "Dynamics of the Electron" 1905) instead of the work due to an unknown internal energy.[27] Therefore, Kaufmann again concluded that his measuring procedures were not compatible with the elementary hypothesis posited by Lorentz-Einstein. [28]

---

[24] Kaufmann, Walter, "Über Die elektromagnetische Masse des Elektrons", *Physikalische Zeitschrift* 4, 1902, p. 56.
[25] Cushing, James, "Electromagnetic Mass, Relativity, and the Kaufmann Experiments", *American Journal of Physics* 49, 1981, pp. 1142, 1148.
[26] Kaufmann, Walter, "Über die Konstitution des Elektrons", *Annalen der Physik* 19, 1906, pp. 487-553.
[27] In 1905 Poincaré examined the three laws – Abraham's, Lorentz's, and "the hypothesis of Langevin" (equivalent to Bücherer's), and claimed in his paper, "On the Dynamics of the Electron", that Abraham's rigid spherical electron was at odds with the principle of relativity. The Lorentzian contracted or deformable electron was the only one to be compatible with the principle of relativity.

Kaufmann concluded, from 1905 onwards, that the mathematical expression proposed by Alfred Bücherer could also be in accord with his measurements and that one could not definitively decide between that expression and that of Abraham as it was derived from his experiments. In the same paper, Kaufmann noted that the two theories of Lorentz and Einstein yielded the same equations of motion for the electron, and he gave the first clear account of the basic theoretical difference between Lorentz's and Einstein's views. [29]

In the annual general meeting of the German Society of Scientists and Physicists (Deutsche Gesellschaft der Naturforscher und Ärrzte) in Stuttgart, on the 19[th] of September 1906, scientists discussed three world pictures, the electromagnetic theories of Abraham, Bücherer, or the other picture based on Lorentz and Einstein's "Principle of Relativity". A discussion revolving around the foundations of physics was held after Planck's lecture. The participants in the discussion were, among others, Kaufmann, Planck, Bücherer, Abraham, Arnold Sommerfeld and others. Scientists did not yet distinguish between Lorentz's theory and Einstein's theory. There were two main theories relating to the electron: Abraham's and Lorentz-Einstein's. An inclination towards Einstein and Lorentz's theories, on the part of scientists such as Planck and Max Laue, was evident.

Einstein did not participate the Stuttgart annual meeting. Einstein was still sitting in the Patent office as an expert II class, and did not participate in the Kaufmann discussion, because he was absent from the Stuttgart meeting. Planck sent Einstein a report of Kaufmann's results and of the ensuing discussion, but added that this was not the updated results. Planck wrote Einstein on November 9, 1907: "In response to your request, I am sending you by the same mail my 'Postscript to the Discussion of Kaufmann's Deflection Measurements' together with the 'Discussion' itself. But I would like to add immediately that Herr Kaufmann subsequently carried out a calculation of the influence that ions produced by the β-rays exert on the electrical field between the condenser plates, from which it follows that the electrical field is extraordinary close to being homogeneous. This calculation of Kaufmann's will

---

One would gain a possible explanation for the contraction of the electron, in supposing that the deformed electron was subject to some kind of constant external pressure (the "Poincaré pressure"), the work done by it being proportional to the variations in the electron's volume. Therefore, one was led to propose an additional external potential, yielding non- electromagnetic external forces, the Poincaré pressure that would stabilize Lorentz's electron while in motion. Poincaré analyzed the mechanism of the contraction of Lorentz's electron and the configuration of the contracted electron. If we consider the configuration of the spherical electron at rest, as the result of electrostatic internal repulsion plus an external pressure caused by the external ether, then we can look at the equilibrium configuration of the moving electron. This configuration is the one that minimizes the total potential energy of the superposed actions on the electron: electrostatic repulsion and Poincaré's pressure. Poincaré, Henri, "Sur la dynamique de l'électron", *Rendiconti del Circolo Matematico di Palermo* 21, 1906, pp. 129-175; pp. 151-166.

[28] Kaufmann, 1906, p. 553.
[29] *CPAE*, Vol. 2, "Einstein on the Theory of Relativity", note 80, p. 267; Kaufmann, 1906.

appear very soon in the *Berliner der Physikalischen Gesellschaft*".[30] Kaufmann repeated his experiments and validated his results.

Einstein *immediately* commented on the situation in a review article in 1907 that he submitted a few weeks later to Johannes Stark's *Jahrbuch der Radioaktivität und Elektonik*. On November 1st, 1907 Einstein wrote Stark "I have now finished the first part of the work for your *Jahrbuch*; "I am working diligently on the second [part] in my, unfortunately rather scarce, free time". The first part dealt with the special theory of relativity. Once Einstein obtained Planck's letter on November 9, 1907, he sat and added a new section dealing with Kaufmann's results. Einstein estimated that the whole paper would be 40 printed pages long, and he told Stark that he hoped he would send him the manuscript "by the end of this month".[31] The paper was published on December 4, 1907.[32]

Einstein like Planck was skeptical and he wrote in the paper: "Only after a more diverse body of observations becomes available will it be possible to decide with confidence whether the systematic deviations are due to a not yet recognized source of errors, or to the circumstance that the foundations of the theory of relativity do not correspond to the facts".[33]

Einstein added: "It also should be mentioned that Abraham's and Bücherer's theories of the motion of the electron yield curves that are significantly closer to the observed curve than the curve obtained from the theory of relativity. However, the probability that their theories are correct is rather small, in my opinion, because their basic assumptions concerning the dimensions of the moving electron are not suggested by theoretical systems that encompass larger complexes of phenomena".[34]

In the 1906 discussion following Planck's lecture, it focused on Planck's idea, which demonstrated that Kaufmann's results had indicated the need for a rapprochement to the principle of relativity, as well as towards Abraham or Bücherer's models, which were not based on the principle of relativity. Planck re-examined Kaufmann's experiments and data analysis and did not find anything seriously amiss in Kaufmann's interpretation of his data. Nevertheless, Planck believed that Kaufmann's data was not a definitive verification of Abraham's theory or a refutation of Lorentz's.

The first participant in the discussion to comment on Planck's idea was Kaufmann himself. Kaufmann thought that it would be best if he were the one to comment on Planck's suggestion, because he had performed the experiments, and therefore he felt he was in a position to evaluate Planck's attempts, "concerning the conclusions it follows from the facts of the follow-up that neither Lorentz's theory nor Abraham's theory agree with them. This is a clear conclusion. Lorentz's theory is even less in accordance than Abraham's theory. The deviations of Lorentz's theory [...] are so great that nowhere is it possible to explain them by follow-up mistakes. Following that, as far as no principled mistake has occurred in the follow-up, Lorentz's theory is invalidated."[35]

---

[30] Planck to Einstein, November 9, 1907, *CPAE*, Vol. 5, Doc. 64.
[31] Einstein to Stark, November 1, 1907, *CPAE*, Vol. 5, Doc. 63.
[32] Including a part on gravitation.
[33] Einstein, 1907a, p. 439.
[34] Einstein, 1907a, p. 439.

To this Planck replied, "We would be able to approach Lorentz's theory closer than Abraham's theory. Depending on the fact that the deviations in one theory are fewer than in the other, we could not determine which one was preferable".[36]

At the 1906 meeting Planck spoke of *Relativtheorie*, "relative theory". Planck demarcated between two models: "Abraham's, according to which, the electron has the shape of a rigid sphere, and the Lorentz-Einstein's, according to which the 'Principle of Relativity' obtains precise validity. For abbreviation, I will refer to the first theory as a 'Kugeltheorie', and call the second 'Relativtheorie'".[37] In the discussion afterwards this soon became *Relativitätstheorie* "relativity theory". In his arguments with other physicists and his comments on their work, Einstein was progressively, and reluctantly, drawn into this new terminology, though in headings and in the text of his own publications he continued to speak of the "relativity principle".

**4.2 Bücherer's experiments**

In the 1906 discussion following Planck's lecture, Bücherer was probably already vacillating and he said after Planck,[38]

"I have followed the results and have reached a few conclusions concerning Kaufmann's measurements, which I want to specify here. [...] relying on the theory of relativity we arrive at the conclusion that other forces will operate when the rays of Becquerel are not directed any more in parallel but towards the plates of the condenser. From here follows an easy access to test the principle of relativity theory depending on Maxwell's equations, only by operating Becquerel's rays in inclination towards the electric or magnetic field. In a perpendicular movement surprisingly the same forces are received as with Lorentz. I have already thought that perhaps creating an angle caused the deviation in Kaufmann's measurements".

Bücherer then was almost prepared to give up his model,[39]

"I wish to repeat the lecturer's comment [Planck's] that my theory [of the electron] is not sufficiently developed in order to continue discussing it here. I have engaged in it and in its conclusions very intensively and I have found that it does not give a more significant contribution than the previous and the later theory of Lorentz.[...] the electron theory distorted in fixed volume [Bücherer's electron model] and the distorted system accordingly contributes almost the same contribution as the new Lorentz theory. On expanding Planck's calculations on to my electron I would not be able to say what will follow from these calculations regarding my electron. [...]

---

[35] Planck, Max (1906b) "Die Kaufmannschen Messungen der Ablenkbarkeit von β-Strahlen in ihrer Bedeutung für die Dynamik der Elektronen," *Physikalische Zeitschrift* 7, 1906, pp. 753-761; pp. 759-760.
[36] Planck, 1906b, p. 760.
[37] Planck, 1906b, p. 756.
[38] Planck, 1906b, p. 760.
[39] Planck, 1906b, p. 760.

Certain arguments can be raised against Einstein's theory of relativity. [...] Having understood that all the theories until now, including my theory, do not answer all the requirements, I asked myself whether it was possible to reach an agreement, with the knowledge of preserving Maxwell's equations and on the basis of the principle of the equality of action and reaction".

Indeed after 1906, Bücherer renounced his electron model that had lead to his expression for the variation of mass with velocity, and he began to gravitate towards the Lorentz-Einstein model. In 1908 he conducted experiments on the beta rays of radium that were more precise than those of Kaufmann, and demonstrated that the Lorentz's model (and that of Einstein also) was a better representation of the experimental variation of the mass of the cathode rays with their velocity. Bücherer concluded his paper by stating that the experimental results of his experiments showed that scientists should incline towards the Lorentz-Einstein theory and that "this result is the confirmation of the principle of relativity".[40]

Bücherer wrote Einstein on September 7, 1908, "First of all I would like to take the liberty of informing you that I have proved the validity of the relativity principle beyond any doubt by means of careful experiments. The experimental design might be known to you from my note in the *Physikal. Zeitschrift*." Bücherer explained in the letter to Einstein the way he undertook the test. He then wrote, "I have thereby definitively disproved my own principle".[41]

Bücherer told Einstein, "I think I can rightly claim that I attained substantially greater precision in all measurements than did Kaufmann, so that I was convinced right from the outset that I must obtain definitive results. I am enclosing a photograph of one of my radiograms, from which you will immediately recognize the superiority of my method, i.e., if you have seen Kaufmann's radiograms. Kaufmann has a great many sources of error in his experimental setup, and I have already told him about one of them".[42] Planck's opinions in 1906 might have influenced Bücherer when he conducted his experiments that confirmed Lorentz's theory.

---

[40] Bücherer, Alfred, "Messungen an Becquerelstrahlen. Die experimentelle Bestätigung der Lorentz-Einsteinschen Theorie", *Physikalische Zeitschrift* 9, 1908, p. 760.
[41] Bücherer to Einstein, September 7, 1908, *CPAE*, Vol. 5, Doc. 117.
[42] Bücherer to Einstein September 9, 1908, *CPAE*, Vol. 5, Doc. 119.

## 4.3 Born and the rigid body problem

In 1908, at the time of his return to Göttingen, Max Born was already thinking of the rigid body problem and entered into polemics with Abraham on his model for the mass of the electron, [43]

"I frequently came into contact with Abraham, I was well informed about his controversy with Lorentz. Abraham was anti-relativist and objected to Lorentz's derivation. I also doubted it, but doubted Abraham's derivation as well. Both proceeded by calculating the self-energy of a charged rigid body in uniform motion (Lorentz with contraction, Abraham without it) as a function of velocity and using this energy as Hamilton's function for obtaining the equations of motion. This procedure assumes that the energy calculated for constant velocity also holds for accelerated motion. My doubts were concerned with this point, and I decided to derive the equations of motion for an accelerated electron in strict accord with the principle of relativity.

This led at once to a great difficulty. For if a body is accelerated, different points of it have different velocities, hence different contractions: the idea of rigidity breaks down. My first problem was therefore: how far can the concept of a rigid body be preserved in relativity? Rigidity means lack of deformation. I worked out a mathematical expression for the deformation of a moving body on relativistic principles, the so-called strain-components, which are differential expressions containing the derivatives of the coordinates as functions of their initial values and of time. Then I postulated that these strain-components should be zero; I obtained in this way differential equations for the possible strain-free motions and I found a solution of them which represents a uniformly accelerated rigid motion in a straight line, uniform in the same sense that the acceleration in the instantaneous rest-system is constant, and rigid relative to the same system […] The main result was a confirmation of Lorentz's formula for the electromagnetic mass as a function of velocity, and its dependence on acceleration. I worked on this investigation all through the winter and spring of 1909 […] I ventured to apply to the Mathematical Society for permission to give a report on it and was admitted."

And then, "Abraham joined in the debate [after the lecture] to tell me that my knowledge of physics seemed to be just as scantly as that of mathematics. He was annoyed because my theory led to Lorentz's formula for the electromagnetic mass and not to his."[44]

*I wish to thank Prof. John Stachel from the Center for Einstein Studies in Boston University for sitting with me for many hours discussing special relativity and its history.*---

[43] Born, Max, *Mein Leben: Die Erinnerungen des Nobelpreisträgers*, 1975, München: Nymphenburger Verlagshandlung GmbH; *My Life Recollections of a Nobel Laureate*, 1978, New York, Charles Scribner's Sons, p. 134.
[44] Born, 1975/1978, p. 135.